\documentclass[pre,showpacs]{revtex4-1}

\usepackage[pdfpagemode=None,pdfstartview=FitH]{hyperref}
\usepackage{graphicx}
\usepackage{epstopdf}
\usepackage[usenames]{color}

\def\br{{\bf r}}
\def\bu{{\bf u}}
\def\bw{{\bf w}}

\begin{document}

\title{Analysis of microtubule motion due to drag from kinesin walkers} 

\author{J. M. Deutsch}
\author{M. E. Brunner}
\affiliation{Department of Physics, University of California, Santa Cruz CA 95064}
\author{William M. Saxton}
\affiliation{Department of Biology, University of California, Santa Cruz CA 95064}

\begin{abstract}
We analyze the nonlinear waves that propagate on a microtubule that is tethered at its minus end
due to kinesin walking on it,
as is seen during the fluid mixing caused by cytoplasmic streaming in Drosophila oocytes.
The model we use assumes that the microtubule can be modeled as an elastic string in a viscous
medium. The effect of the kinesin is to apply a force tangential to the microtubule and we
also consider the addition of a uniform cytoplasmic velocity field. We show that travelling
wave solutions exist and analyze their properties. There exist scale invariant families
of solutions and solutions can exist that are flat or helical. The relationship between
the period and wavelength is obtained by both analytic and numerical means. Numerical
implementation of the equation of motion verifies our analytical predictions.
\end{abstract}

\maketitle

\section{Motion of Microtubule With Kinesin Walkers}

Here we study the motion of microtubules in the context of cytoplasmic
streaming in stage 10B-11 Drosophila oocytes.  At that stage, the roughly
hemispherical oocyte is bounded by a plasma membrane and an underlying
cortex comprised of an actin filament meshwork.  Long microtubules,
with minus ends attached to the cortex, have their plus ends free.
Kinesin-1 motor proteins that walk along the microtubules toward plus
ends generate opposing forces on the microtubule and the surrounding
viscous cytoplasm.  Cytoplasm is thus moved toward plus ends, generating
vigorous flows for mixing.  Because minus ends are anchored, the equal
but opposite force on a microtubule generates dynamic bending along
its length.  The results of this work have been used to understand this
phenomenon in recent work by the authors~\cite{DeutschBrunnerSaxton}.

We begin with the equation for a microtubule with an applied force tangent to the direction of chain. The position
of the microtubule at arclength $s$ and at time $t$ is denoted $\br(s,t)$. As is usual at small scales, all inertial effects are negligible
and the system is dominated by the drag coefficient per unit length $\nu$,
\begin{equation}
\label{eq:microtubule}
\nu \frac{\partial \br}{\partial t} =  -C \frac{\partial^4 \br}{\partial s^4} + \frac{\partial}{\partial s}(T(s)\frac{\partial \br}{\partial s}) -
f_k \frac{\partial \br}{\partial s} + \nu v_{s}{\hat k} .
\end{equation}
$C$ denotes the elastic bending constant of the microtubule. The tension $T(s)$ enforces the inextensibility of microtubules which
can be written as  $|\partial \br/\partial s| = 1.$ 
Kinesin motors walking along the microtubule are assumed to be carrying cargo that acts as an impeller pushing the surrounding
cytoplasmic medium. By Newton's third law, this force produces a drag on the impellers that transmits this force to the
attached microtubule. Assuming a high density of kinesin walkers, the effective force on the microtubule will be in the
direction of the local tangent to $\br(s)$ which is $\partial \br/\partial s$. The coefficient $f_k$ gives the strength of the force.
Lastly we add a uniform flow field representing the velocity of the cytoplasm $v_s$ in the vicinity of the microtubule
which we take to be in the $\hat k$ direction.

We rescale to dimensionless variables 
\begin{equation}
\label{eq:rescale}
\sigma = s/\rho_0,~ \bu = \br/\rho_0, ~ \tau = t \omega_0,  ~ T' = T \rho_0^2/C, ~{\rm and} ~ h = \nu v_s/f_k 
\end{equation}
so that
\begin{equation}
\label{eq:dimensionless}
\frac{\partial \bu}{\partial \tau} =  - \frac{\partial^4 \bu}{\partial \sigma^4} + \frac{\partial}{\partial \sigma}(T'(\sigma)\frac{\partial \bu}{\partial \sigma}) -
\frac{\partial \bu}{\partial s} + h {\hat k} .
\end{equation}
This requires that 
\begin{equation}
\label{eq:scaling}
\omega_0 = f_k/(\nu \rho_0), ~  {\rm and} ~ \rho_0^3 = C/f_k. 
\end{equation}
$\rho_0$ and $\omega_0$ are constants that make our new variables dimensionless.

We will consider solutions where one end (the {\em minus} end) is tethered to a point, say the origin, so that
$\br(s=0,t) = 0$ and is freely hinged at that point. The other end is free.
The total arclength of the microtubule is denoted by $L$.
The general characteristic of all numerical solutions to Eq. \ref{eq:microtubule} is that
for large $s$ the solutions appear to be travelling waves. In fact, the form of solution
becomes independent of $L$ so in fact we could consider this problem in the limit $L\rightarrow \infty$.
For example we find solutions that asymptotically become helical with a fixed path and radius for
large $s$. Other solutions are planar and are periodic in $s$. In all these cases the solution
has the form of a travelling wave that we analyze in detail below.

\section{Steady State Travelling Wave Solutions}

Now we look for steady state travelling wave solutions. First we clarify what this means.
The whole microtubule should not be translating because one end, (far away from where
the travelling wave solution is valid) is tethered. 
The position averaged over one period of a monomer, $\langle u \rangle$, should be  time independent.
If the microtubule is stretched out
in one particular direction, there should be a displacement $\bu_d$ of the microtubule about a straight
line solution,
\begin{equation}
\label{eq:travelling}
\bu(\sigma, \tau) = \bu_d(\sigma-v\tau) + \alpha \sigma {\hat k}.
\end{equation}
The term $\bu_d$ describes a displacement that travels along the backbone of the chain maintaining its shape.
Therefore we require
\begin{equation}
\label{eq:ud_constraint}
\frac{\partial \langle \bu_d \rangle}{\partial \sigma} = 0
\end{equation}

Substituting Eq. \ref{eq:travelling} into Eq. \ref{eq:dimensionless} gives a left hand side
of $-v \partial \bu_d/\partial \sigma$ and a term of the form $-\partial \bu_d/\partial \sigma$ on
the right hand side. Choosing $v = 1$ and in addition $\alpha = h$, we are left with

\begin{equation}
\label{eq:reduced_steadystate}
\frac{\partial^4 \bu}{\partial \sigma^4} + \frac{\partial}{\partial \sigma}(T'(\sigma)\frac{\partial \bu}{\partial \sigma}) = 0.
\end{equation}
We define $\bw = \partial \bu/\partial \sigma$  and
note that the inextensibility requirement $|\partial \br/\partial s| = 1$ implies that $|\bw| = 1$. We can write 
\begin{equation}
\label{eq:bw=bd_alpha}
\bw  = \frac{\partial \bu_d}{\partial \sigma} + \alpha {\hat k}
\end{equation}
Note that $\bw(\sigma,\tau)$ depends only on $\sigma -\tau$. Because $\bw$ is a function
of only one variable $\xi \equiv \sigma -\tau$, 
we can 
integrate Eq. \ref{eq:reduced_steadystate} with respect to $\xi$ obtaining
\begin{equation}
\label{eq:sphericalmotion}
\frac{d^2 \bw}{d \xi^2} - T'(\sigma)\bw = -g {\hat k}.
\end{equation}
The integration constant on the right hand side $-g {\hat k}$ must lie in the $\hat k$ direction by symmetry.
This is the same equation as that of the classical mechanics of a particle of unit mass, with position $\bw$ travelling on the surface of a 
unit sphere under the influence of gravity, with the variable $\xi$ being analogous to time. The term $-T'$ is a normal
force that constrains the particle to stay on the sphere's surface.

From Eq. \ref{eq:ud_constraint} and Eq. \ref{eq:bw=bd_alpha} we see that the motion is subject to the
constraint
\begin{equation}
\label{eq:alpha_constraint}
\langle \bw \rangle = \alpha {\hat k}
\end{equation}
The constant $g$ and the initial conditions of the particle must be chosen so as to satisfy this constraint.
There is an infinite number of solutions satisfying these conditions leading to different 
steady state solutions for the microtubule.

The spherical pendulum Eq. \ref{eq:sphericalmotion} has been analyzed~\cite{CushmanBates} in detail. In general
the orbits will not be closed, but we will relegate
our discussion below to two cases where this is the case, circular motion in Eq. \ref{eq:sphericalmotion}, corresponding to helical
conformations of a microtubule, and motion in the $x-z$ plane, corresponding to solutions similar in shape to cycloids.

\subsection{Scale Invariant Family of Solutions}
\label{subsec:ScaleInvariance}

One important point to notice about the general structure of Eq. \ref{eq:sphericalmotion} is its invariance under
change of $\xi$, that is $\xi \rightarrow \lambda \xi$. A change in scale of $\lambda$ does not modify the
solution, because $T'(\xi)$  and $g$ are both constraints, and can therefore be chosen arbitrarily. So that if
$\bw(\xi)$ is a solution to Eq. \ref{eq:sphericalmotion}, so is $\bw(\lambda \xi)$. Furthermore all solutions
of this form will have the same $\langle \bw \rangle$ and hence the same $\alpha$. Note that a length scale
shift of $\bw = \bw_0(\lambda \xi)$ corresponds to a displacement of $\bu = (1/\lambda)\bu_0(\lambda \xi)$, which
changes scale because of the $1/\lambda$ prefactor in this expression. This means that for every
steady state conformation of the microtubule, there exists a family of steady state solutions with identical
shape but arbitrary size.

\subsection{Circular Orbits}
\label{subsec:CircularOrbits}

Circular orbits are solutions to Eq. \ref{eq:sphericalmotion}. Eq. \ref{eq:alpha_constraint} requires that the vertical
height of the orbit be at $w_z = \alpha$. Therefore the radius perpendicular to $\hat k$ (the $w_x, w_y$ plane) is 
\begin{equation}
\label{eq:w_perp_alpha}
w_\perp \equiv \sqrt{1-\alpha^2}. 
\end{equation}
It is first easiest to analyze this by direct analogy to the classical mechanical problem of a unit mass particle moving on a unit sphere (the
spherical pendulum).
If the height above the sphere is $\alpha$, then application of Newton's laws gives that the tension must balance the force
of gravity $g$ and the centripetal acceleration $v^2/w_\perp$ giving
\begin{equation}
\frac{g}{v^2} \equiv \frac{\alpha}{1-\alpha^2} 
\end{equation}
Where $v$ is not the real velocity of the microtubule but $|d \bw_\perp/d\xi|$. This is the rate at which the
tangent vector rotates in the $x-y$ plane. 
Because $|d\bu_d/d\xi|  = |\bw_d| = r_c$ is a constant, $\bw_d$ is a circle. 
Therefore according to Eq. \ref{eq:travelling}, this will give a helical conformation of the microtubule.

In the spherical pendulum analogy, the angular velocity of the particle
is $\Omega_p = v/w_\perp$. The total arclength of the chain (in dimensionless units) of a single period of
the chain is
\begin{equation}
\label{eq:SRalpha}
S = \frac{2\pi R}{\sqrt{1-\alpha^2}}
\end{equation}
and $S \Omega_p = 2 \pi$ so that $S = 2\pi/\Omega_p = w_\perp/v$. Comparing this with Eq. \ref{eq:SRalpha}
we obtain 
\begin{equation}
R = \frac{w_\perp}{v} \sqrt{1-\alpha^2}. 
\end{equation}
Using Eq. \ref{eq:w_perp_alpha} this gives 
\begin{equation}
\label{eq:Ralpha_v}
R = \frac{1-\alpha^2}{|\frac{d w_\perp}{d \xi}|} = \frac{1-\alpha^2}{v}
\end{equation}
Because $g$ is a free parameter, $v$ can be any positive number and therefore Eq. \ref{eq:Ralpha_v}
implies that the radius of the helix can be arbitrary. 

Because the velocity of this travelling wave is unity, the spatial and temporal periodicities are equal. 
Therefore the period of rotation of the helix $P'$ is equal to  $S$ in Eq. \ref{eq:SRalpha}, in the dimensionless units that we are using.
Therefore the relationship between the period and the pitch is
\begin{equation}
\label{eq:Pdimensionless}
P' = \frac{2\pi R}{\sqrt{1-\alpha^2}}
\end{equation}
In terms of our original units the period is
\begin{equation}
\label{eq:T_circle}
P = \frac{2\pi R \nu}{f_k \sqrt{1-\alpha^2}}
\end{equation}

\subsection{Two Dimensional Solutions}
\label{subsec:2dsolns}

\begin{figure}[htp]
\begin{center}
\includegraphics[width=\hsize]{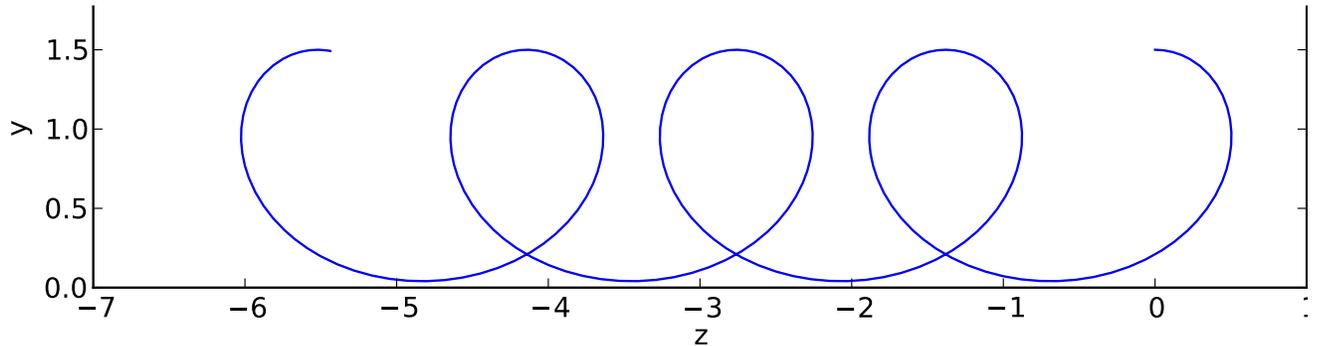}
\caption{ 
A two dimensional travelling wave solution for a microtubule that looks close to a prolate cycloid but
differs from it in functional form.
}
\label{fig:prolate}
\end{center}
\end{figure}

We now consider orbits in the $x-z$ plane. This corresponds to a pendulum swinging around vertically through the bottom of the sphere.
This is a standard introductory mechanics problem. Letting $\theta$ denote the angle with respect to the $-{\hat k}$ direction,
\begin{equation}
\label{eq:pendulum}
\ddot{\theta} = -g \sin\theta 
\end{equation}
where the dots that normally represent a time derivative are really derivatives with respect to $\xi$. Low energy solutions
where the total energy $E$ is less than $g$ give $\theta$ oscillating between two bounds. 
In this case the curve for the microtubule will look sinusoidal. Beyond the separatrix where $E > g$, $\theta$ will
increase without bound. This gives rise to microtubule conformations that look close to prolate cycloids. In other words,
the microtubule forms a travelling wave that periodically loops backwards. This is shown in Fig. \ref{fig:prolate}

Small values of $|\alpha|$ correspond to low values of $g/E$. In this case, the particle spends almost
equal times at all points on the circle giving a small value of $\langle w\rangle = \alpha$. As $g/E$
increases to $1$, $\alpha$ increases and the loops become tighter; a situation that is very costly energetically.
However this is not the only solution to these equations for a fixed value of $\alpha$. If instead we
consider solutions with $E < g$, then large $g/E$ corresponds to small oscillations of $\bw$ about $-{\hat k}$.
In this case, $|\alpha|$ is close to $1$ and the curve has no tight loops. The shape of the microtubule
is close to a sine wave.

To obtain the relationship between $G \equiv g/E$ and $\alpha$ we use conservation of energy
to obtain the period $T$,
\begin{equation}
\label{eq:pendulumenergy}
\dot{\theta} = \sqrt{2E(1+G\cos\theta)}
\end{equation}

\begin{figure}[htp]
\begin{center}
\includegraphics[width=\hsize]{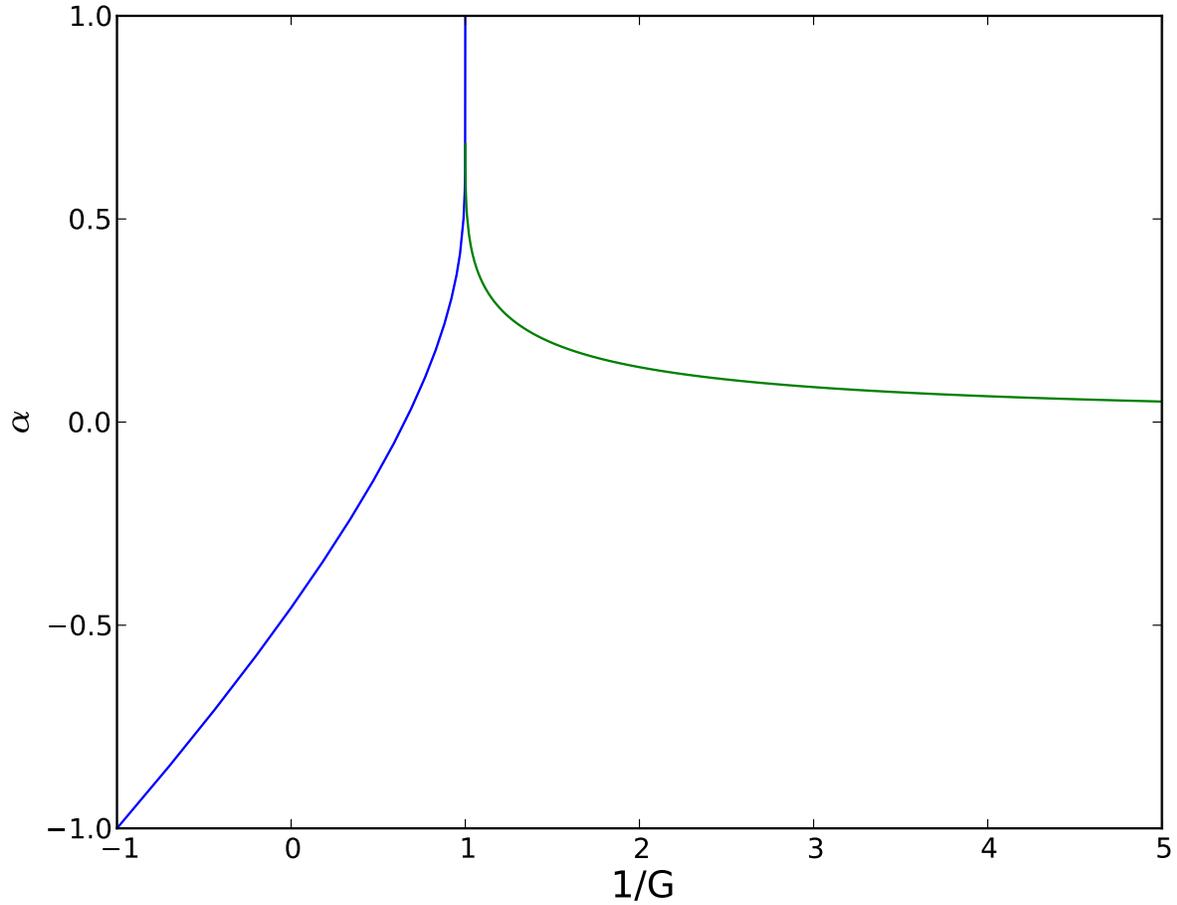}
\caption{ 
A plot of the value of $\alpha$ as a function of $1/G$.
}
\label{fig:AlphaVsE}
\end{center}
\end{figure}

The time average of $\bw$ in the $\hat k$ direction is
\begin{equation}
\label{eq:avependulumcostheta}
\langle \cos\theta\rangle = \frac{\int_0^\pi \frac{\cos\theta}{\sqrt{1+G\cos\theta)}} d\theta}{\int_0^\pi \frac{1}{\sqrt{1+G\cos\theta)}} d\theta}
\end{equation}
This can be expressed in terms of elliptic integrals as
\begin{equation}
\label{eq:avecosthetaElliptic}
\langle \cos\theta\rangle =  \frac{1}{G} (1- (1+G) \frac{{\cal E}(2G/(1+G))}{{\cal K}(2G/(1+G))})
\end{equation}
Where  $\cal E$ and $\cal K$ are the complete Elliptic integrals of the first and second kind respectively.
For bounded orbits, below the separatrix, the corresponding expression can be calculated giving
\begin{equation}
\label{eq:avecosthetaEllipticBounded}
\langle \cos\theta\rangle =  -2 \frac{{\cal E}(\frac{e+1}{2})}{{\cal K} (\frac{e+1}{2}) }) + 1
\end{equation}
A plot of the $\alpha = \langle \cos\theta\rangle$ versus $1/G = E/g$ is shown in Fig. \ref{fig:AlphaVsE}
using these two expressions.
$\alpha \rightarrow 0$ in the limit as $G \rightarrow 0$. Note that there are two possible values of
$E/g$ corresponding to one value of positive $\alpha$.

The case of $\alpha = 0$
corresponds to the case of circular orbits discussed in Sec.  \ref{subsec:CircularOrbits}. 
The cytoplasmic streaming velocity is $\propto h$ (see Eq.  \ref{eq:dimensionless}).
When $h = 0$, we have shown that the solutions are circular. If $h$ is now made very
small, we expect that the solution will only be slightly perturbed from circular
solutions. This corresponds to a small value of $g/E$. Therefore the value of
$G$ that is chosen for small enough $h$ should correspond to the larger value of
$1/G$ shown in Fig.  \ref{fig:AlphaVsE}. This corresponds to almost circular loops
shifting slightly forward after every turn. As $h$ increases the loops become
tighter giving rise to a high elastic bending energy. At some point therefore, 
we expect a transition to another state. In fact, simulations discussed in Sec. \ref{subsec:CompWithSims} show that near a
flat surface, the microtubule transitions out of the plane to a helical shape 
at $h \approx .385$ for $128$ links in a chain. For
still larger $h$, the microtubule becomes flat again looking close to sinusoidal.

\section{Numerical Implementation of the Full Solutions}

By comparing the steady state solutions found above for large arclength $s$ and  time $t$, we will
see that these solutions are physically realistic ones to consider. As we will see, the full
equations of motions go to solutions of this type. However this is not a complete
description of the problem, because of these solutions
a particular one is selected from a whole family of solutions. The same behavior occurs
in other problems in pattern selection such as dendritic growth~\cite{Kessler,Barbieri}.

\begin{figure}[t]
\begin{center}
\includegraphics[width=\hsize]{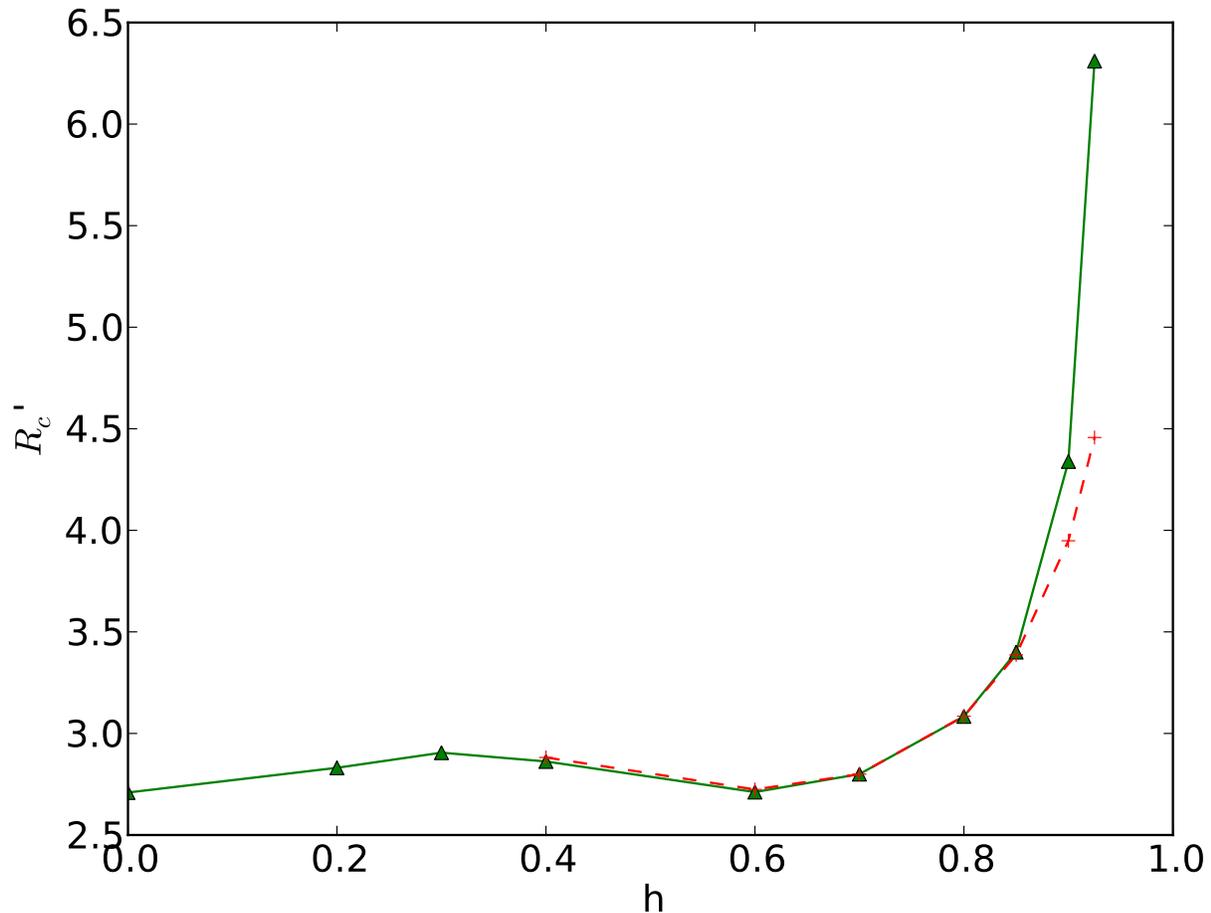}
\caption{ 
The rescaled radius of curvature $R_c/\rho_0$ versus the rescaled velocity field $h=\alpha$.
The green triangles are for $64$ link chains, the red $+$ symbols are for chains of $128$ links.
}
\label{fig:RadCurvVSh}
\end{center}
\end{figure}

\subsection{Comparison With Simulation}
\label{subsec:CompWithSims}

Eq.  \ref{eq:microtubule} was analyzed numerically using a method similar to that
used earlier in the context of gel electrophoresis~\cite{DeutschElectrophoresisScience,DeutschMadden}.
Link length drift was handled using a similar procedure to that implemented for chains with inertia~\cite{DeutschCerfFriction}
One end was constrained to have coordinates at the origin, while the other was free. 

A Runge Kutta time step was $0.001$, $C = 20$, $f_k = 2$ and simulations were performed with $N=64$ or $N=128$ links,
as will be noted below.

\begin{figure}[htp]
\begin{center}
(a) \includegraphics[width=0.45\hsize]{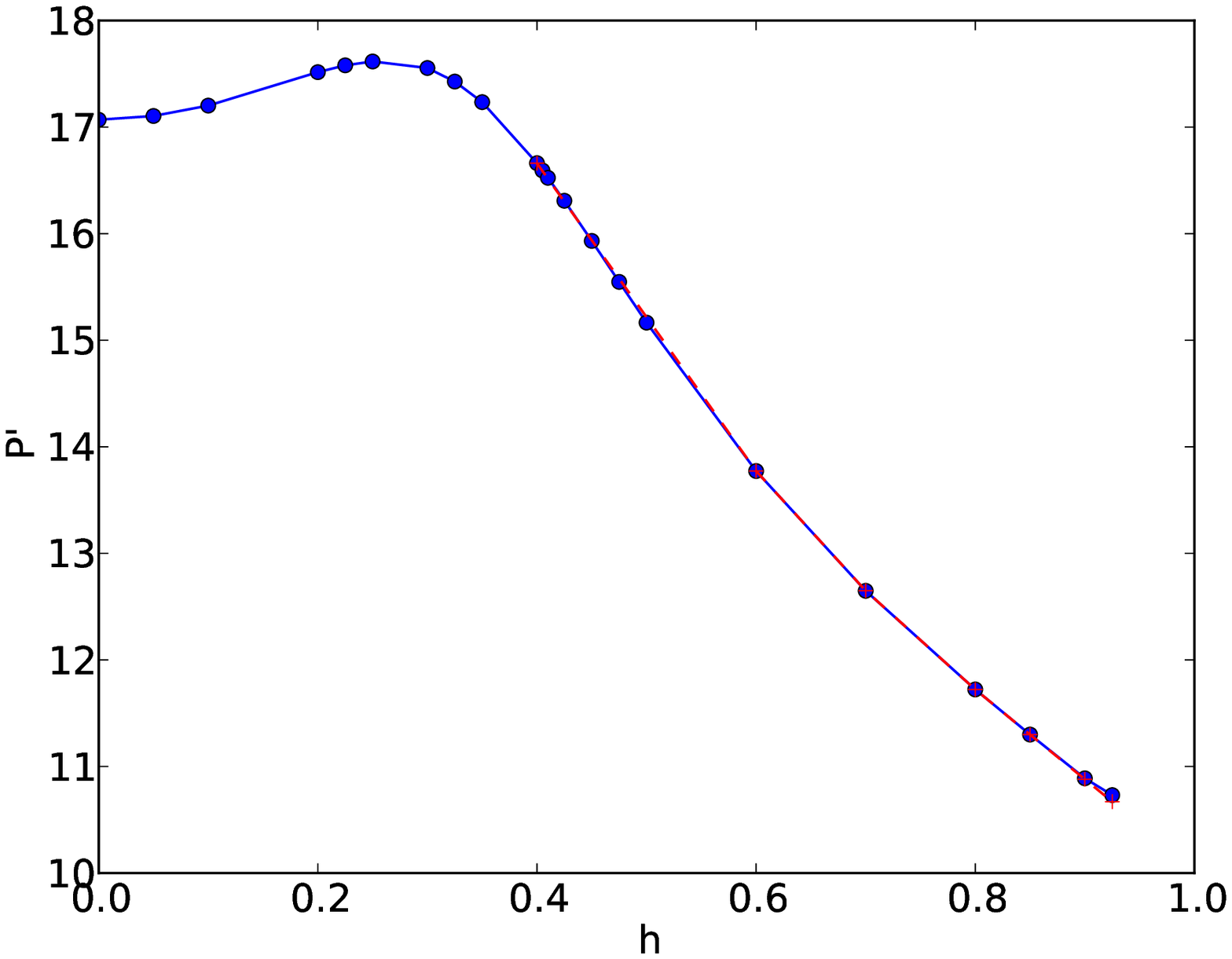}
(b) \includegraphics[width=0.45\hsize]{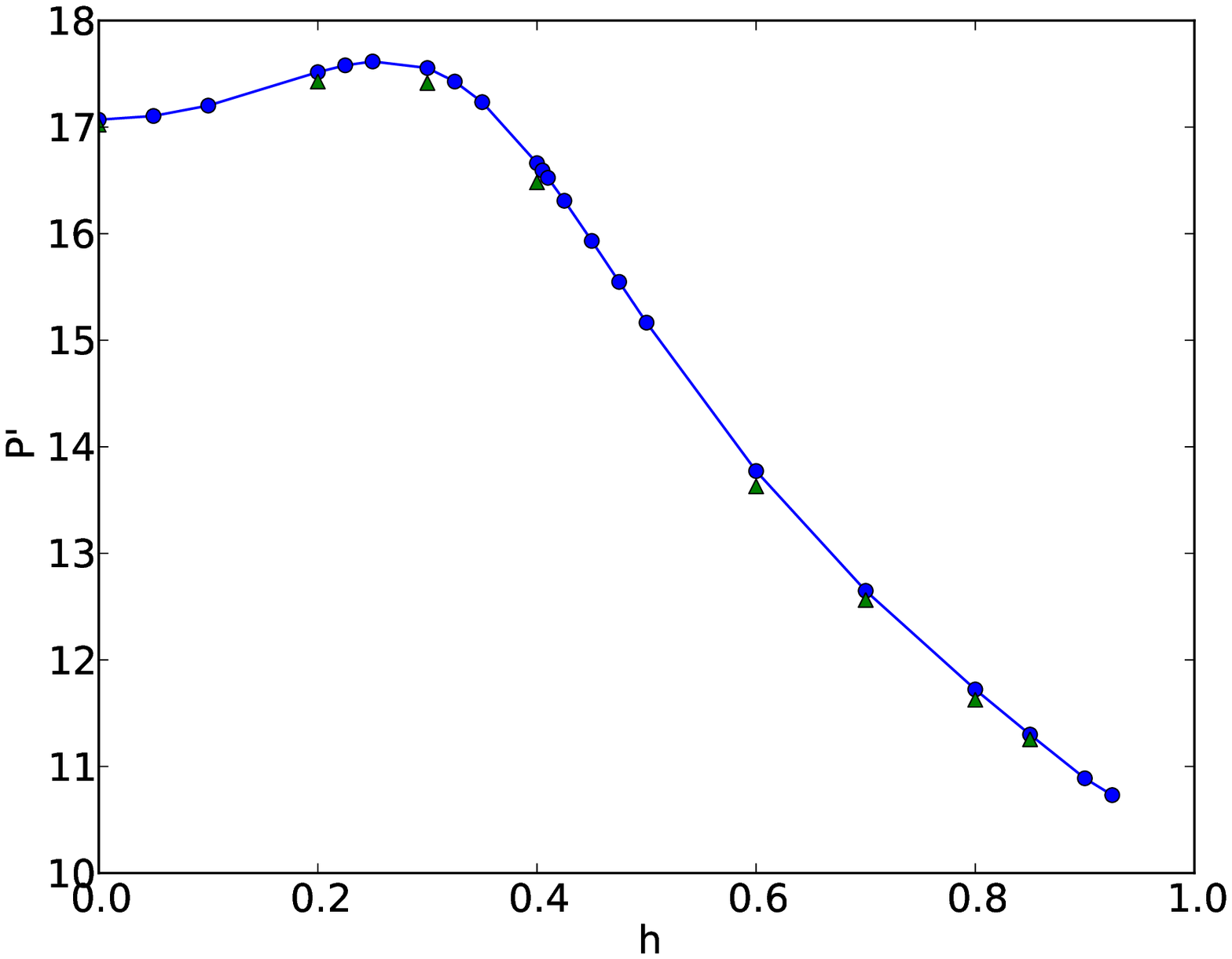}
\caption{ 
(a)The rescaled period $P' = P\omega$ versus the rescaled velocity field $h=\alpha$.
The blue circles are for $64$ link chains, the red $+$ symbols are for chains of $128$ links.
(b)
The rescaled period $P' = P\omega$ versus the rescaled velocity field $h=\alpha$
calculated by using Eq. \ref{eq:PvsRc} and data on the radius of the helix versus $h$ shown
in Fig. \ref{fig:RadCurvVSh}. The result is shown by the green triangles. 
The blue circles are the same data shown in Fig. \ref{fig:PeriodVSh} by directly
measuring the period. 
}
\label{fig:PeriodVSh}
\end{center}
\end{figure}

We first consider the case of a microtubule with no wall or other external forces aside from the external velocity field.
As a function of the rescaled external velocity field $h$, the radius of curvature and period were calculated once the
system had reached steady state. Using rescaled variables as defined by Eqs. \ref{eq:rescale} and \ref{eq:scaling} we
can write the dimensionless radius of curvature as $R_c' = R_c/\rho_0$, and the dimensionless period as $P' = \omega P$.
The radius of curvature was calculated at the middle of the chain to reduce finite size effects.
The results are shown in Figs.  \ref{fig:RadCurvVSh} and \ref{fig:PeriodVSh}. The data for $R_c$, Fig. \ref{fig:RadCurvVSh} using
$64$ links, is very close to those of $128$ links except for the highest $h$ values. The data for the period, Fig. \ref{fig:PeriodVSh}(a)
show good agreement for both chain sizes at all values of $h$ studied.

We now test the analytic predictions that we made relating the period to the radius of the helix and $\alpha$
given by Eq. \ref{eq:Pdimensionless}, and our conclusion that $\alpha = h$. The relationship between the
radius of curvature $R_c$ and the radius $R$ of the helix is readily calculated to be $R = R_c (1-\alpha^2)$.
Therefore Eq. \ref{eq:Pdimensionless} gives
\begin{equation}
\label{eq:PvsRc}
P' = 2 \pi R_c' \sqrt{1-\alpha^2}
\end{equation}
With this prediction and using the data for $R_c$ in Fig. \ref{fig:RadCurvVSh}, we can independently calculate $P'$ for $64$
link chains. We can only do this where finite size effects are not important and therefore omit the highest two $h$ values
from this analysis. The data, Fig. \ref{fig:PeriodVSh}(b), show excellent agreement to within the differences expected by finite size effects. This
corroborates our analytical analysis of steady state solutions. 

The case of $h = \alpha = 0$ displayed in Fig.  \ref{fig:RadCurvVSh} can be written in terms of the original
dimensional variables of Eq. \ref{eq:microtubule} using Eq.  \ref{eq:scaling}
\begin{equation}
\label{eq:R}
R_c =  (C/(\beta f_k))^{1/3}.
\end{equation}
From the numerical solution, this gives $\beta = 0.05 \pm 0.0005$. This is useful in analyzing experimental data, as
$R_c$ is nearly constant for $h < 0.7$.

\begin{figure}[htp]
\begin{center}
\includegraphics[width=\hsize]{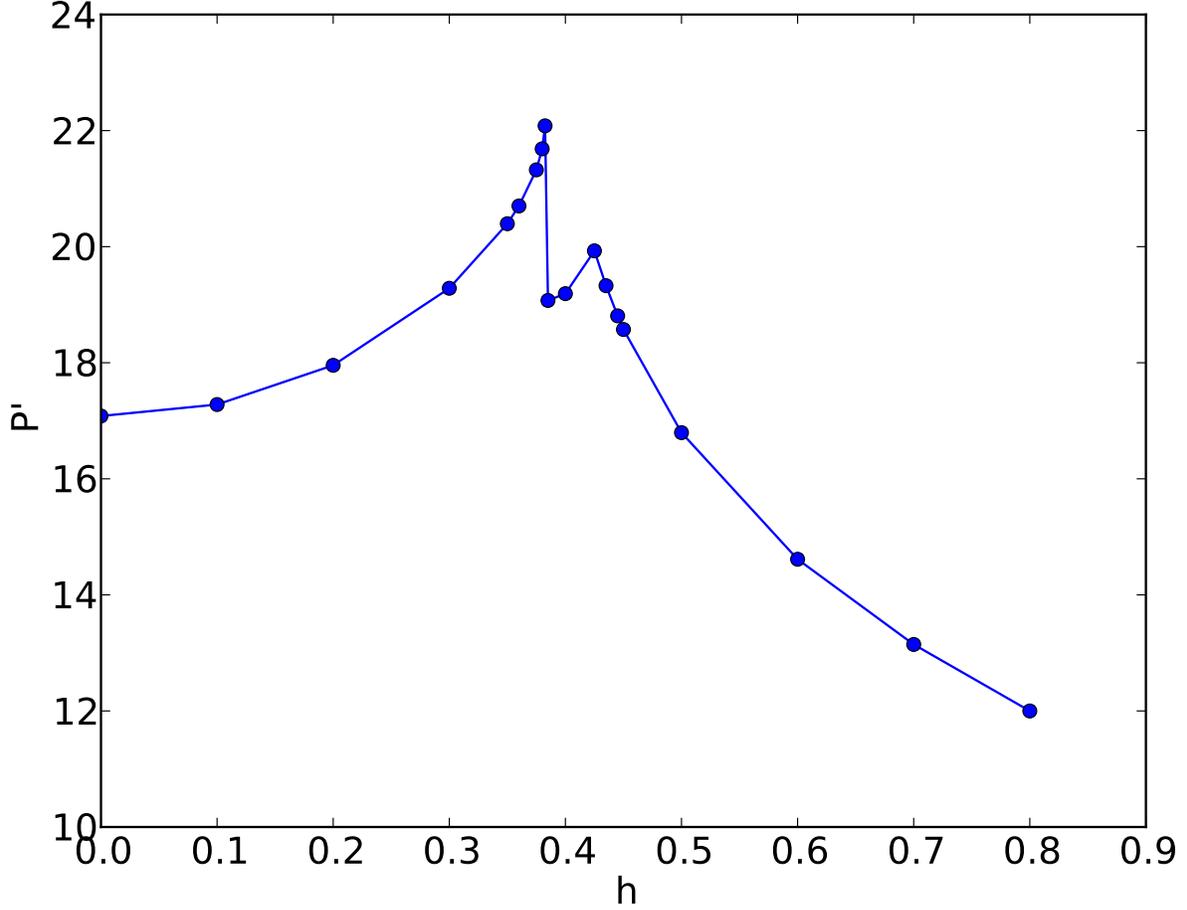}
\caption{ 
The rescaled period $P' = P\omega$ versus the rescaled velocity field $h=\alpha$
for chains of $128$ links.
}
\label{fig:WallPeriodVSh}
\end{center}
\end{figure}

Next we consider the presence of a wall. We introduced a force representing a wall in the $y-z$ plane of the form
\begin{equation}
\label{eq:wallforce}
{\bf f}_w =  \frac{x^2}{(x+1)^2} { \hat  i}  
\end{equation}
which is only present for $x < 0$. The force is singular at $x=-1$ preventing the chain from crossing that plane.
The period as a function of $h$ is shown in Fig. \ref{fig:WallPeriodVSh} for $C=20$ and chains with $128$ links.

For small values of $h$, the period does not vary much. The shapes obtained are flat and
appear to be the same as those found in \ref{subsec:2dsolns}. However there is non-analytic behavior at $h \approx 0.385$
corresponding to a transition to flattened helical waves. Then at $h \approx 0.425$ the solution becomes almost
completely flat and showing waves that look closer to sine waves, corresponding to the flat 
sinusoidal-like solutions studied in \ref{subsec:2dsolns}.

\begin{figure}[htp]
\begin{center}
(a)
\includegraphics[width=0.45\hsize]{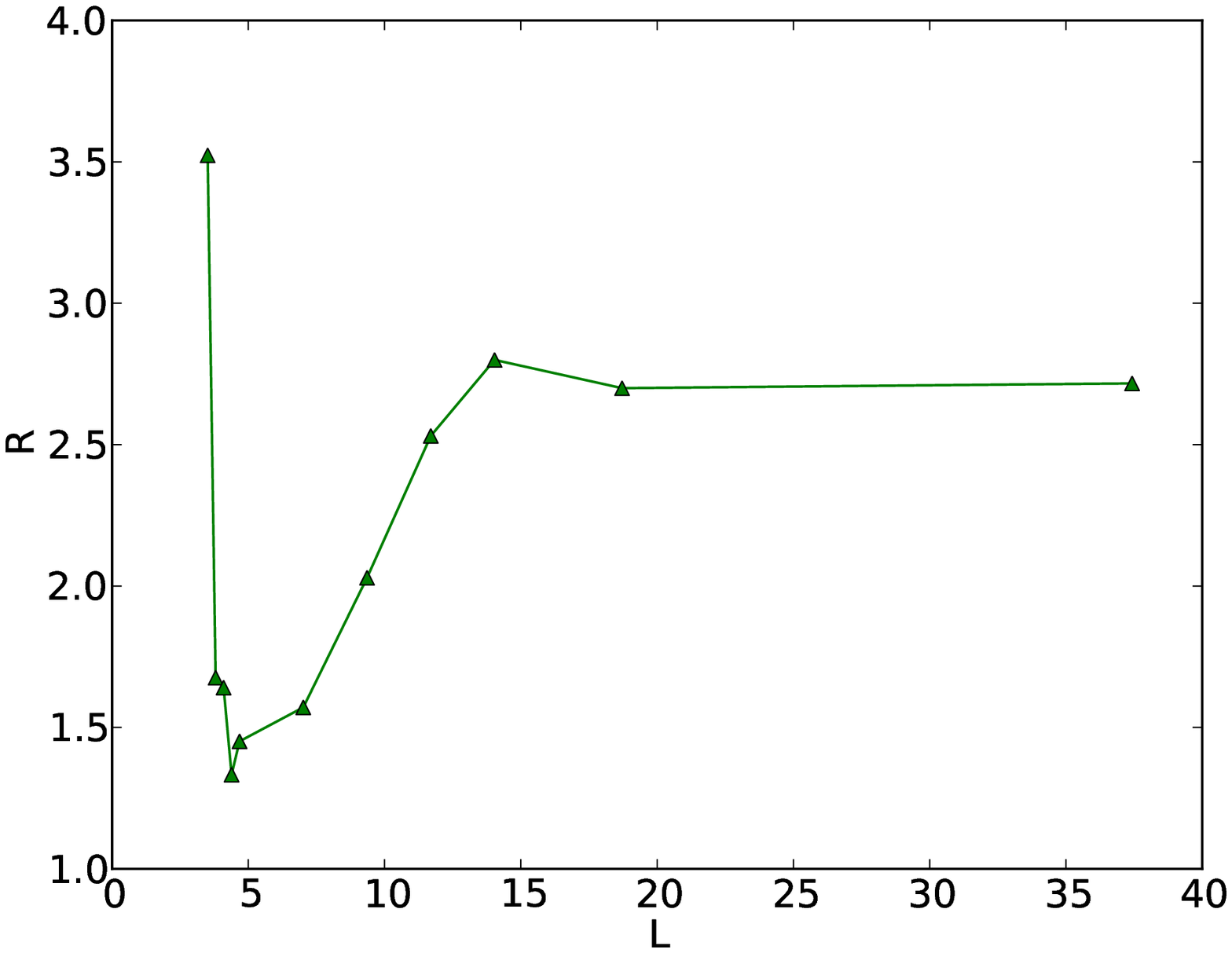}
(b)
\includegraphics[width=0.45\hsize]{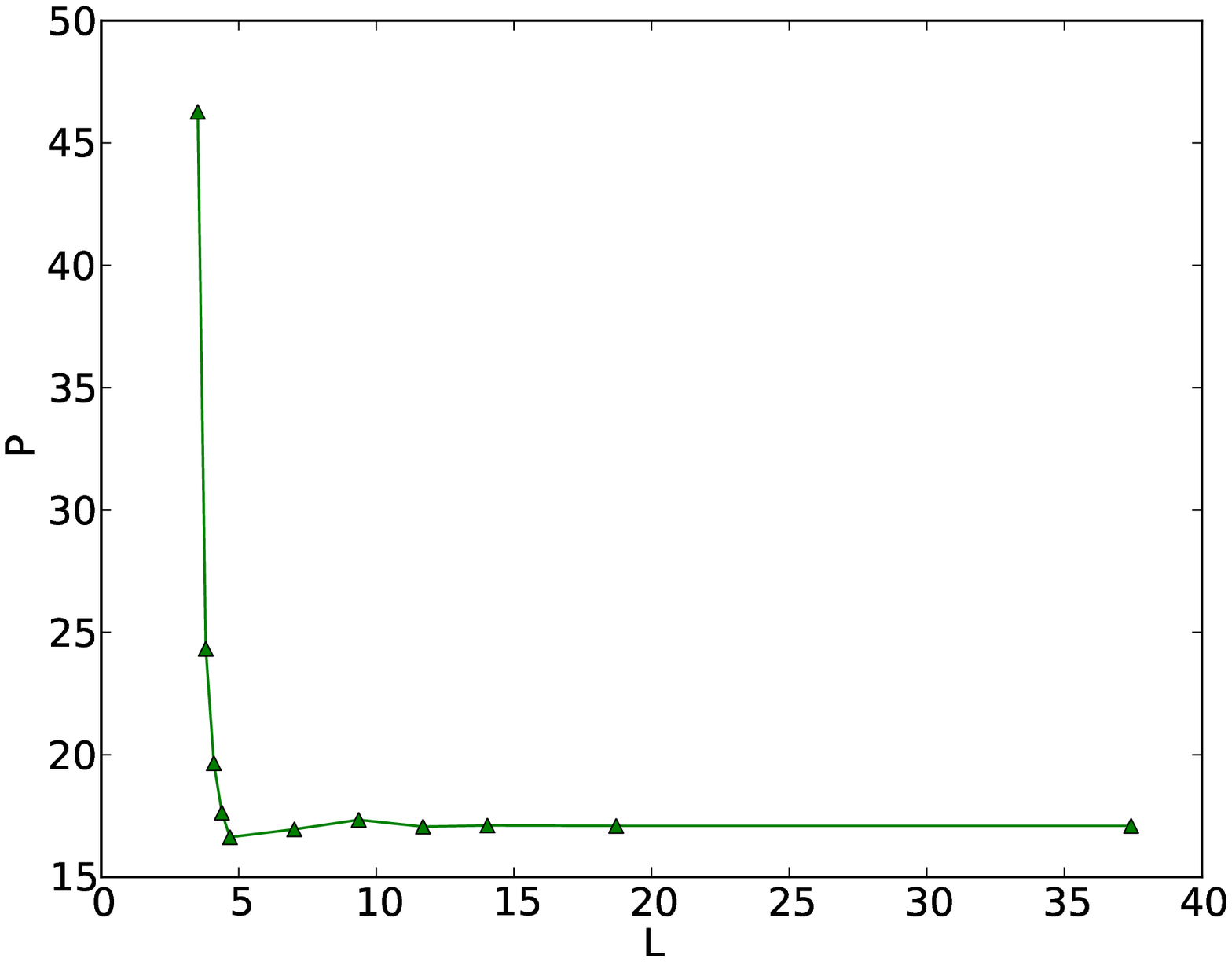}
\caption{ 
(a)
The rescaled radius of curvature measured at the middle link of a chain  versus the rescaled chain length in steady state for $h=0$.
(b)
The rescaled period  versus the rescaled chain length in steady state for $h=0$.
}
\label{fig:RvsL}
\end{center}
\end{figure}

We now consider the dependence of period and curvature on the length of chains $L$. Our
analytical solutions have been in the limit that the chain length $L\rightarrow \infty$.
For finite length chains, we expect corrections to this behavior. This is important
to analyze in relationship to experiments using microtubule gliding assays~\cite{ValeSchnappEtAl}. In this case
we consider $h=0$ and study how the rotating spiral wave solutions vary with increasing
length. Fig. \ref{fig:RvsL}(a) shows the dependence of the radius of curvature measured
halfway along the arclength, as a function of chain length. As usual, the rescaled dimensionless
variables, see Eqs. \ref{eq:rescale} and \ref{eq:scaling} have been used. The curve
shows a non-monotonic dependence on $L$ that levels off at $L \approx 20$. 
Fig. \ref{fig:RvsL}(b) shows that the rescaled period is slightly non-monotonic as well but decreases to a constant
value at $L \approx 15$. In comparison with gliding assays, we took~\cite{DeutschBrunnerSaxton} $L = 14.0$, where
values of parameters are within $3\%$ of their asymptotic values.

Note that in Fig. \ref{fig:RvsL}(b) the period diverges for finite $L \approx 3$. Below this
point, the solution is a static straight line. This is similar to the usual buckling transition
for a finite length elastic rod. The microtubule needs to be sufficiently long to undergo
the dynamical instability analyzed here.

\subsection{Selection of Scale}

As noted in section \ref{subsec:ScaleInvariance}, there exists
a continuous family of steady state solutions because  if
$\bw(\xi)$ is a solution, then so is $\bw(\lambda \xi)$ for arbitrary
$\lambda$. The general way that a particular value of $\lambda$ is selected
is likely to be due to the same mechanism as in other pattern formation
problems~\cite{Kessler,Barbieri}. The travelling wave solutions ignore the
boundary conditions at the ends $s=0$ and $s=L$.  Far from those ends,
we have found a continuous family of solutions. However these solutions
become invalid close to the ends.  The travelling wave solutions are
only valid in the limit of $0 << s << L$. The full solution must match
to one of these travelling solutions in that region but will differ
greatly near the ends.

As an example, consider the circular solutions of
Sec. \ref{subsec:CircularOrbits} with $h = \alpha = v_s = 0$. There the
steady state solution is a rotating circle wrapped around on itself of
{\em arbitrary radius} $R$.  However near the end $s=0$, the solutions
goes into the circle's center. Similarly it deviates from a circle
at $s=L$.  In analogy with other pattern growth problems such as the
``Geometric Model"~\cite{Kessler},  where the mathematics have been analyzed
in detail, we expect that the boundary conditions imposed on the ends,
will only be satisfied for certain values of $R$. If there is more than
one allowed value of $R$, the value picked out will be the most stable of
these. In the Geometric Model the form of the equations is much simpler,
making it possible to understand the overall structure of the problem
more easily. In the present case, a precise understanding of the numerical
results will be the subject of future research.

\section{Conclusions}

We have analyzed a model that describes a microtubule tethered at its minus
end and subject to viscous drag and forces due to kinesin walking toward the plus end. The walkers
provide a force tangent to the local direction of the microtubule and
cause an instability that causes nonlinear waves to travel along its
arclength. The scale of the wave can be arbitrary. An open question now 
is to find an analytic model that predicts the precise scale that
is selected. There are many other questions that are interesting.
What is the motion when there are obstacles that prevent free motion
of the microtubule? Do these lead to chaotic behavior? What happens
in the case of many microtubules interacting through steric and hydrodynamic
interactions? This last question is difficult to answer and will
be tied into the question of large scale collective motion of forests of
microtubules and the velocity flows that they produce. Velocity
flows giving rise to micro-mixing have been studied in related systems~\cite{GoldsteinTuvalvandeMeentPNAS,GoldsteinTuvalvandeMeentPRL,MeentSedermanGladdenGoldstein,VerchotLubiczGoldstein} and it would be very interesting to
understand the large scale cytoplasmic motion for Drosophila oocytes.

\section{Acknowledgements}

The authors would like to thank Ian Carbone, and Bill Sullivan for useful discussions.
This material is based upon work supported by National Institutes of Health GM046295 (to W.M.S.), and National Science Foundation
CCLI Grant DUE-0942207 (to J.M.D.).

{\em Note:} After the completion of this work, we came across work of Bourdieu
et al.~\cite{Bourdieu95} which analyzed their experimental results on
spirals in  mysosin and kinesin gliding assays using the two dimensional
version of Eq. \ref{eq:microtubule}, obtaining the identical scaling,
and simulating it using a nearly identical approach. This provides
further evidence that the mechanism we have proposed for cytoplasmic
streaming is physically viable.

\end{document}